\documentclass{iopjournal}


\fancyhfoffset[L]{0mm}
\fancyhfoffset[R]{0mm}

\usepackage{graphicx}
\usepackage{booktabs}
\usepackage{dcolumn}
\usepackage{bm}
\usepackage{multirow}
\usepackage{soul}
\usepackage{lineno}
\usepackage{amsmath}
\usepackage{amssymb}
\usepackage{url}
\usepackage{hyperref}
\hypersetup{colorlinks=true,linkcolor=blue, linktocpage}

\begin{document}


\title{Overlap-aware segmentation for topological reconstruction of obscured objects}


\author{J.~Schueler$^{1,*}$, H.~M.~Ara\'ujo$^{2,3}$, S.~N.~Balashov$^3$, J.~E.~Borg$^4$, C.~Brew$^3$, F.~M.~Brunbauer$^5$, C.~Cazzaniga$^6$, A.~Cottle$^7$, D.~Edgeman$^1$, C.~D.~Frost$^6$, F.~Garcia$^9$, D.~Hunt$^8$, M.~Kastriotou$^6$, P.~Knights$^{10}$, H.~Kraus$^8$, A.~Lindote$^{11}$, M.~Lisowska$^{5}$, D.~Loomba$^1$, E.~Lopez~Asamar$^{12}$, P.~A.~Majewski$^{3,2}$, T.~Marley$^{2}$, C.~McCabe$^{13}$, L.~Millins$^{10,3}$, R.~Nandakumar$^3$, T.~Neep$^{10}$, F.~Neves$^{11}$, K.~Nikolopoulos$^{10,14}$, E.~Oliveri$^5$,
A.~Roy$^2$,
T.~J.~Sumner$^2$, E.~Tilly$^{1}$, W.~Thompson$^1$,
M.A.~Vogiatzi$^{2,3}$}

\affil{$^1$Department of Physics and Astronomy, University of New Mexico, Albuquerque, NM, 87131, USA}

\affil{$^2$Department of Physics, Blackett Laboratory, Imperial College London, London, SW7 2AZ, UK}

\affil{$^3$Particle Physics Department, STFC Rutherford Appleton Laboratory, Didcot, OX11 0QX, UK}

\affil{$^4$Luleå University of Technology, 97187 Luleå, Sweden}

\affil{$^5$CERN, 1211 Geneva 23, Switzerland}

\affil{$^6$ISIS Neutron and Muon Source, STFC Rutherford Appleton Laboratory, Didcot, OX11 0QX, UK}

\affil{$^7$University College London (UCL), Department of Physics and Astronomy, London WC1E 6BT, UK}

\affil{$^8$Department of Physics, Keble Road, University of Oxford, Oxford, OX1 3RH, UK}

\affil{$^9$Helsinki Institute of Physics, University of Helsinki, FI-00014 Helsinki, Finland}


\affil{$^{10}$School of Physics and Astronomy, University of Birmingham, Birmingham, B15 2TT, UK}


\affil{$^{11}$LIP -- Laborat\'{o}rio de Instrumenta\c{c}\~{a}o e F\'{\i}sica Experimental de Part\'{\i}culas, University of Coimbra, P-3004-516 Coimbra, Portugal}


\affil{$^{12}$Departamento de Fisica Teorica, Universidad Autonoma de Madrid, 28049 Madrid, Spain}

\affil{$^{13}$Department of Physics, King’s College London, London, WC2R 2LS, UK}

\affil{$^{14}$University of Hamburg, 22767, Hamburg, Germany}


\affil{$^*$Author to whom any correspondence should be addressed.}

\email{jschueler1@unm.edu}


\begin{abstract}
The separation of overlapping objects presents a significant challenge in scientific imaging. While deep learning segmentation-regression algorithms can predict pixel-wise intensities, they typically treat all regions equally rather than prioritizing overlap regions where attribution is most ambiguous. Recent advances in instance segmentation show that weighting regions of pixel overlap in training can improve segmentation boundary predictions in regions of overlap, but this idea has not yet been extended to segmentation regression. We address this with Overlap-Aware Segmentation of ImageS (OASIS): a new segmentation-regression framework with a weighted loss function designed to prioritize regions of object-overlap during training, enabling extraction of pixel intensities and topological features from heavily obscured objects. We demonstrate OASIS in the context of the MIGDAL experiment, which aims to directly image the Migdal effect--a rare process where electron emission is induced by nuclear scattering--in a low-pressure optical time projection chamber. This setting poses an extreme test case, as the target for reconstruction is a faint electron recoil track which is often heavily-buried within the order(s)-of-magnitude brighter nuclear recoil track. Compared to unweighted segmentation regression, we demonstrate OASIS's novel overlap region-targeted loss function weight to be the single most important training weight for improving intensity and topological reconstructions of the low-energy electron tracks that tend to be most dominated by pixel overlap. Averaging over eight training campaigns, we further show the addition of overlap-targeted weights to improve median intensity reconstruction errors from $-41.1\%$ to $-13.3\%$ for these low-energy electrons. These performance gains demonstrate OASIS as a generalizable methodology for recovering obscured signals in overlap-dominated regions. All code is openly available to facilitate cross-domain adoption.
\end{abstract}


\section{Introduction}
\label{sec:intro}

A pervasive challenge in scientific imaging is the decoupling of overlapping objects. Some examples include the blending of galaxies in crowded astronomical fields where overlapping light profiles can bias photometry and redshift estimations$\,$\cite{SCARLET}; fluorescence microscopy, where multiple fluorophores can overlap within a diffraction-limited PSF$\,$\cite{speiser2021deep}; diverse applications in medical imaging \cite{2019arXiv190208128Z,2023arXiv230314373J,2025arXiv250109116L}; and particle detectors, where overlapping particle tracks can obscure features of interest such as decay vertices or faint secondaries$\,$\cite{Alonso-Monsalve:2023xgh,DUNE:2025wti}. 
In these and other scientific imaging applications where pixels (or voxels) represent physically meaningful quantities, we aim to demonstrate that prioritizing regions of overlap in the training objective of pixel-intensity segmentation-regression models provides a path toward disentangling overlapping objects and improves the attribution of object-specific intensities even in regions where intensities of overlapping pixels span orders of magnitude.

Deep learning-based segmentation networks provide pixel-level reconstructions that have become essential tools across scientific imaging. Architectures such as U–Net$\,$\cite{ronneberger2015unet} and Mask R-CNN$\,$\cite{2017arXiv170306870H} have enabled object-level separation in diverse domains, from biomedical imaging to astronomy, by mapping each pixel to a categorical label or probabilistic mask. Extending these architectures to segmentation regression allows networks to predict continuous, physically meaningful quantities such as voxel-wise parameter maps in quantitative medical imaging$\,$\cite{Torrado-Carvajaljnumed.118.209288} and flux maps in galaxy deblending$\,$\cite{2019MNRAS.490.3952B,2024PASA...41...35Z}. These models jointly reconstruct object topology and pixel-wise intensities, however they treat all regions equally rather than prioritizing overlap regions where attribution is most ambiguous. Occlusion-aware instance-segmentation frameworks such as MultiStar$\,$\cite{2020arXiv201113228W} and BCNet$\,$\cite{2021arXiv210312340K}, on the other hand, explicitly emphasize regions of pixel-overlap in their training, but neither regress physical intensities. Here we introduce a novel framework called Overlap-Aware Segmentation of ImageS (OASIS) that extends overlap-aware training to segmentation regression. Specifically, OASIS employs programmable weights in its loss function that target both objects and regions. Object-targeted weights prioritize pixel-wise intensity attribution for specific object classes and are relatively standard in computer vision. Region-targeted weights, however, are novel and can be tuned to specifically target regions of pixel overlap, allowing training to prioritize intensity attribution in the most ambiguous pixels. The combination of overlap-aware training and segmentation-regression is well suited for high resolution particle detectors like gas time projection chambers (TPCs), where detector images can contain overlapping ionization signals with physically meaningful intensities.

Gas TPCs are widely used in nuclear and particle physics, where their high granularity imaging capability enables reconstruction of energy deposition along particle trajectories. Example applications in which overlapping track signals arise, and where component-wise segmentation-regression could therefore be valuable include neutrinoless double beta decay searches like NEXT$\,$\cite{NEXT:2012zwy,NEXT:2020jmz,NEXT:2021vzd}, optical TPC studies of exotic nuclear decay processes$\,$\cite{2007_Miernik,miernik1,PhysRevC.90.014311}, liquid argon TPC experiments such as MicroBooNE~\cite{MicroBooNE:2016pwy}, which already employs segmentation in its reconstruction chain $\,$\cite{MicroBooNE:2020yze,MicroBooNE:2020hho,MicroBooNE:2021pvo,Grizzi:2025olp,Aurisano:2024uvd}, and searches for the Migdal effect in nuclear scattering by the MIGDAL and MARVEL experiments$\,$\cite{Araujo:2022wjh,Yi:2026fmf}. From a reconstruction standpoint, searches for the Migdal effect are a particularly challenging case because the target signal is an image containing a bright nuclear recoil (NR) track that heavily overlaps with an order(s)-of-magnitude fainter electron recoil (ER) track. In order to measure the Migdal effect and meaningfully compare observations with theoretical predictions, reconstruction of the energy of the faint and heavily obscured ER track is essential, while reconstruction of its emission angle provides an additional test of theory. In this work, we focus on the MIGDAL experiment, whose high-resolution imaging readout and strongly overlapping NR–ER topology provide a direct and challenging testbed to demonstrate the performance and potential of OASIS.

We structure the rest of this work as follows: In Section~\ref{sec:method} we detail the OASIS framework in a general context, including the backbone network we use (Section~\ref{subsec:network}) and its loss function (Section~\ref{subsec:loss}). We then introduce the MIGDAL experiment (Section~\ref{subsec:migdal}) and describe the details relevant for applying OASIS to images recorded by the MIGDAL experiment. In Section~\ref{sec:results}, we present the results of systematic ablation studies comparing intensity and topological reconstruction of overlap-dominated ER tracks across 36 trained OASIS models with distinct training-weight configurations. We then use the best performing model from this study to further quantify OASIS's intensity and angular reconstruction performance. Finally in Section~\ref{sec:conclusion}, we contextualize our results in terms of their impact to the MIGDAL experiment and beyond.

\section{The OASIS Framework}
\label{sec:method}
The general objective of OASIS is to separate an input image consisting of $k$ object classes into $k$ object-specific intensity maps. OASIS's training is guided by a custom loss function that applies object and region-specific weights that are predefined for the application at hand. 
In the case of separating overlapping objects, different weights can be prescribed to differing regions of overlap. For example, if a task were to consist of three unique object-classes, labeled 1, 2, and 3, the four unique cases of object overlap ($1\cap 2$, $1\cap 3$, $2\cap 3$, and $1\cap 2\cap 3$) could each be assigned their own overlap region weights. 

Figure~\ref{fig:unet_schematic} shows a schematic of the OASIS framework. An input $n_x\times n_y$ intensity-map image containing $k$ distinct object-classes is processed through a backbone network\footnote{We employ a standard U--Net backbone in this work, however other suitable backbone networks could be applied.}, which we describe in Section~\ref{subsec:network}. The hierarchical features extracted by the backbone network are then passed into a segmentation-regression head that produces $k$ output intensity maps of the same dimensions, where each output map represents the predicted intensity contribution from one object class.

We train the network using a custom weighted loss function $\mathcal{L}$, described in Section~\ref{subsec:loss}, which can be tuned to assign higher penalties to reconstruction errors in overlap regions (white patches in Figure~\ref{fig:unet_schematic}), as well as regions and channels corresponding to specific objects. Loss function weights that disproportionately penalize misattribution of pixels within regions of object-overlap effectively prioritize reconstructing objects in regions where intensity attribution is most challenging.

\subsection{Network architecture}
\label{subsec:network}
\begin{figure*}
\includegraphics[width=\linewidth]{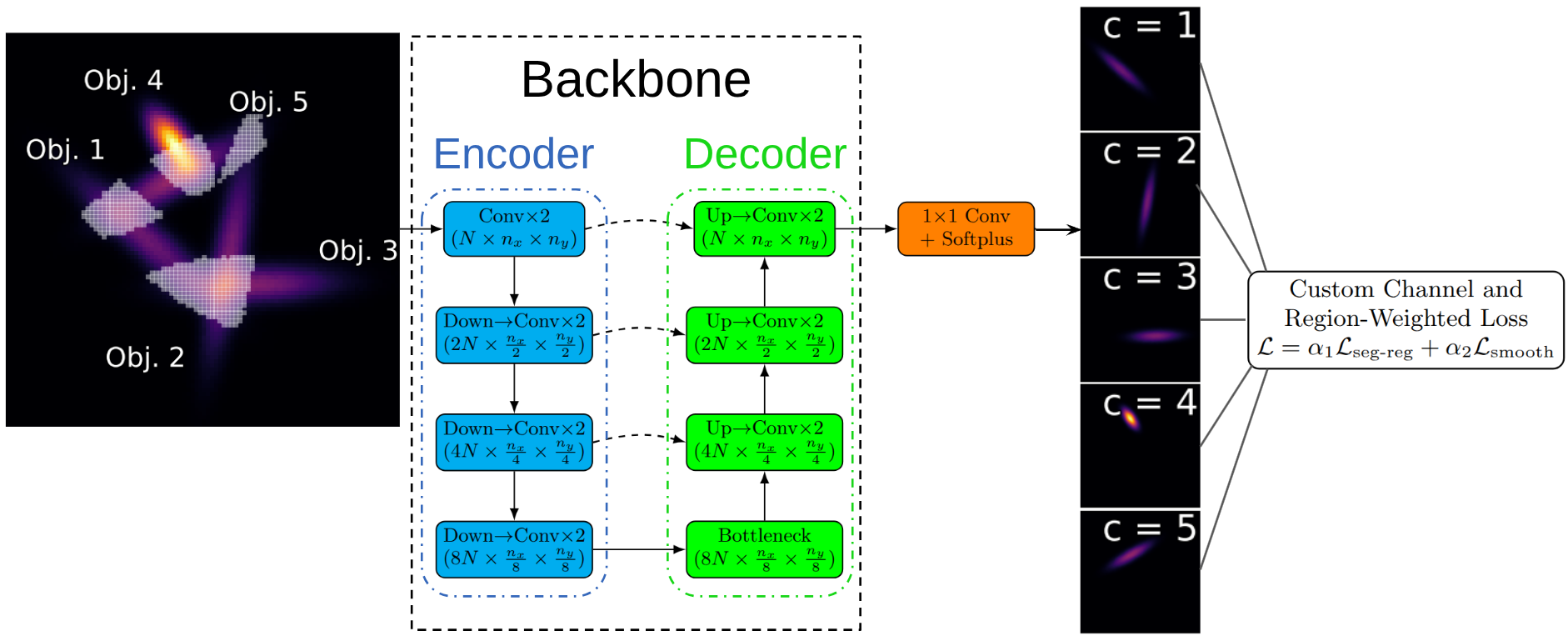}
\caption{Schematic of OASIS. An input intensity-map image of dimension $n_x\times n_y$ containing $k$ distinct object classes ($k=5$ in this example) is passed as input into the network. The backbone network, where here we use U--Net, processes the images which are then passed into a segmentation-regression head, shown as the orange block. This maps the backbone network's output to $k$ intensity maps of dimension $n_x\times n_y$, corresponding to each distinct object class. During training, OASIS is optimized using a custom loss function that compares the predicted output intensity maps with truth. The loss function incorporates weights that can be tuned to assign higher penalties to reconstruction errors in overlap regions (white patches in the input image) and to channels with fainter objects.\label{fig:unet_schematic}}
\end{figure*}

The U--Net we employ consists of a four stage encoder path, a bottleneck layer, and a decoder path that mirrors the encoder path. Skip connections link the corresponding encoder and decoder stages as diagrammed in Figure~\ref{fig:unet_schematic} and explained below. 

Given a single-channel\footnote{In general the number of input channels need not be one, so multi-band images in astronomy, for example, could be used as input to OASIS.} intensity-map input image, $\mathcal{I}(x,y)\in\mathbb{R}^{n_x\times n_y}$, the first encoder block applies two consecutive $N$-channel $3\times 3$ convolutions, each followed by a group normalization over 8 groups and a SiLU~\cite{silu} activation. We refer to this combination of operations as a \textit{standard two-layer block} with an output depth of $N$-channels. The output of this first encoder block is an $N$-channel feature map at full spatial resolution, $\hat{\mathcal{I}}_\mathrm{enc1}\in\mathbb{R}^{N\times n_x\times n_y}$. Each subsequent encoder block begins with a depthwise $3\times 3$ convolution with stride 2 that halves the spatial resolution while maintaining the channel depth, followed by a standard two-layer block that doubles the output depth. After the final encoder block, the output feature map $\hat{\mathcal{I}}_\mathrm{enc4}\in\mathbb{R}^{8N\times \frac{n_x}{8}\times \frac{n_y}{8}}$ is passed into the bottleneck block, which refines this output through another standard two-layer block that retains the same channel depth (8$N$). This output is then passed into the decoder path which mirrors the encoder path. Each decoder block performs the following steps
\begin{enumerate}
\item A bilinear interpolation that doubles the spatial resolution.
\item A $1\times 1$ convolution that halves the channel depth.
\item Apply the skip connection: concatenate this output with the output feature map from the mirrored encoder block. This step doubles the number of channels.
\item Standard two-layer block that halves the output channel depth.
\end{enumerate}
All together, each subsequent decoder block doubles the spatial resolution of the feature maps while halving their channel depth. The decoder path output $\hat{\mathcal{I}}_\mathrm{dec4}\in\mathbb{R}^{N\times n_x\times n_y}$ is then passed through the segmentation head which is a $1\times 1$ convolution with Softplus activation that projects the channel depth to the number of object-classes, $k$. The final output of our network is $\hat{\mathcal{I}}=[\hat{\mathcal{I}}_{1},\ldots,\hat{\mathcal{I}}_{k}]^\top\in\mathbb{R}^{k\times n_x\times n_y}$.
When applying OASIS to the MIGDAL experiment (Section~\ref{subsec:migdal}) we use $512\times 288$ images, a base channel depth of $N=32$, and $k=2$ object-classes.

\subsection{Loss Function}
\label{subsec:loss}
OASIS's novel loss function consists of a custom-weighted pixel-intensity regression loss term and a smoothness regularization term
\begin{equation}
\label{eq:loss}
\mathcal{L}
= \alpha_{1} \, \mathcal{L}_{\text{seg-reg}}
+ \alpha_{2} \, \mathcal{L}_{\mathrm{smooth}},
\end{equation}
where $\alpha_{1}$ and $\alpha_2$ are weights assigned to the two loss terms. To enhance training performance, input images, $\mathcal{I}_\mathrm{in}(x,y)$, can optionally be masked by a pre-defined bounding box region $B\subseteq \mathcal{I}_\mathrm{in}(x,y)$ that isolates regions of interest in the input data. Without loss of generality, we write our loss terms in terms of this bounding box region, $B$, and note that not applying a bounding box mask is equivalent to setting $B = \mathcal{I}_\mathrm{in}$. Descriptions of the two loss function terms follow:
\paragraph{Channel-specific pixel and region-weighted regression loss, $\mathcal{L}_\text{seg-reg}$.}  
This term is responsible for OASIS's segmentation-regression predictions and represents a channel and region-weighted mean absolute error minimization between channel predictions $\hat{\mathcal{I}}_{\mathrm{c}}$ and ground truth $\mathcal{I}_{\mathrm{c}}$, with extra weight given to predefined channels of interest and overlap regions of interest. Given a set of $k$ unique object-classes, this loss aims to simultaneously optimize the channel-wise decomposition of predicted pixel positions and intensities for each object-class:
\begin{align}
\mathcal{L}_{\text{seg-reg}}
= \frac{
\sum_{c=1}^k w_{c} \sum_{(x,y)\in B}
W_{\mathrm{region}}(x,y)\, \big|\hat{\mathcal{I}}_{c}(x,y) - \mathcal{I}_{c}(x,y)\big|
}{
\sum_{c=1}^k w_{c} \sum_{(x,y)\in B} W_{\mathrm{region}}(x,y)
}
\end{align}
Here $w_{c}$ are channel (object)-specific weights, and $W_\mathrm{region}(x,y)$ are region weights that can, for example, be applied to regions where truth input maps from specific channels have nonzero pixel intersection.

\paragraph{Smoothness loss, $\mathcal{L}_\text{smooth}$.}  
This is a standard regularization term$\,$\cite{2016arXiv160309599E} that penalizes the model if neighboring pixel predictions vary by too much. Since this term is designed to be small but nonzero, its weighting factor $\alpha_2$ should be small so that this term doesn't dominate the overall loss:
\begin{align}
\mathcal{L}_{\mathrm{smooth}}
&= \sum_{c=1}^k \sum_{(x,y)\in B}
\Big( |\hat{\mathcal{I}}_c(x+1,y) - \hat{\mathcal{I}}_c(x,y)| + |\hat{\mathcal{I}}_c(x,y+1) - \hat{\mathcal{I}}_c(x,y)| \Big).
\end{align}

The specific choice of loss function weights, preprocessing strategies, training procedures, and evaluation metrics depends on the characteristics of the application domain. While the OASIS framework is general and can be applied to any imaging problem requiring decomposition of any number of overlapping sources into constituent intensity maps, the number of tunable loss function weights in OASIS scales as $n+2^n-1$, where $n$ is the number of target object classes to reconstruct. For problems involving the separation of many object classes, the large number of tunable weights presents a practical limitation in our ability to systematically study the effects different combinations of loss function weights have on performance. This partly motivates our choice of Migdal effect searches as the application used to demonstrate OASIS. In addition to presenting a genuinely difficult reconstruction problem, where a faint ER track must be separated from an order(s)-of-magnitude brighter and heavily overlapping NR track, the problem contains only two objects of interest. As a result, OASIS has only five adjustable loss function weights in this application.

\section{Application of OASIS to the MIGDAL experiment}
\label{subsec:migdal}

Here we describe our application of OASIS to the MIGDAL experiment. We begin by introducing the Migdal effect and the MIGDAL experiment, and then provide parameters relevant to OASIS that are specific to its application to the MIGDAL experiment.

The Migdal effect occurs when the scattering-induced sudden displacement of a nucleus induces the emission of an atomic electron. First theorized by A. Migdal in 1939$\,$\cite{migdal1939ionizatsiya}, this rare effect in quantum mechanics gained renewed attention when Ibe et.$\,$al.$\,$\cite{Ibe:2017yqa} connected it to dark matter searches by tabulating probabilities of electron emission for several atomic species. Since then, numerous experiments have used the Migdal effect to extend their sensitivity to light dark matter$\,$\cite{LUX:2018akb,EDELWEISS:2019vjv,CDEX:2019hzn,XENON:2019zpr,SENSEI:2020dpa,COSINE-100:2021poy,EDELWEISS:2022ktt,DarkSide:2022dhx,XMASS:2022tkr,LZ:2023poo,PandaX:2023xgl,SuperCDMS:2023sql,SENSEI:2023zdf,DAMIC-M:2025luv}, and the effect was observed for the first time in 2026$\,$\cite{Yi:2026fmf}.

The MIGDAL experiment$\,$\cite{Araujo:2022wjh} aims to measure and characterize the Migdal effect using fast neutrons from a deuterium-deuterium generator incident on an optical time projection chamber filled with 50$\,$Torr CF$_4$ gas. Images are recorded with a Hamamatsu ORCA Quest qCMOS camera$\,$\cite{orca}, which captures $2048\times 1152$ images of the scintillation light produced by neutron scattering events that are amplified by two glass gas electron multipliers (GEMs)$\,$\cite{Takahashi:2013rea}. Images are acquired in continuous rolling shutter mode at a rate of 120 frames per second and view an $8.0\times 4.5\,\mathrm{cm}^2$ region through an EHD‐25085-C F0.85 lens$\,$\cite{ehdlens}. At nominal operating GEM gains, the 16-bit pixels of the CMOS sensor provide the dynamic range necessary to simultaneously image and reconstruct both sub-4$\,$keV ERs and up to $\mathcal{O}(\mathrm{MeV})$-scale fast-neutron induced NRs without saturating. In this work, we apply OASIS to both real nuclear recoil (NR) tracks and hybrid simulated images where simulated electron recoil (ER) tracks are stitched onto real NRs, mimicking the Migdal effect signal. Further details of the MIGDAL experiment, detector, and optical camera readout system are given in Refs.$\,$\cite{Araujo:2022wjh,Knights:2024wjh,MIGDAL:2024alc}.

In the region of interest for the MIGDAL experiment's search (ER energies between $\sim$4$\,$keV and $\sim$15$\,$keV), the Migdal effect cross section increases exponentially with decreasing ER energy$\,$\cite{Cox:2022ekg}. This poses a challenge for searches in low pressure electron drift gas mixtures for two reasons. First, the length of ER tracks decreases with energy, and second, diffusion is relatively high in such gas mixtures. It is therefore common for ER tracks to be almost entirely obscured by the diffuse boundary of the accompanying NR track. In these cases it is extremely challenging to detect the ER signal, let alone extract sufficient information to reliably reconstruct its energy and angle. Measurements of the ER energy are necessary to validate theoretical models, so overcoming these limitations is critical.

\begin{figure}
\includegraphics[width=\linewidth]{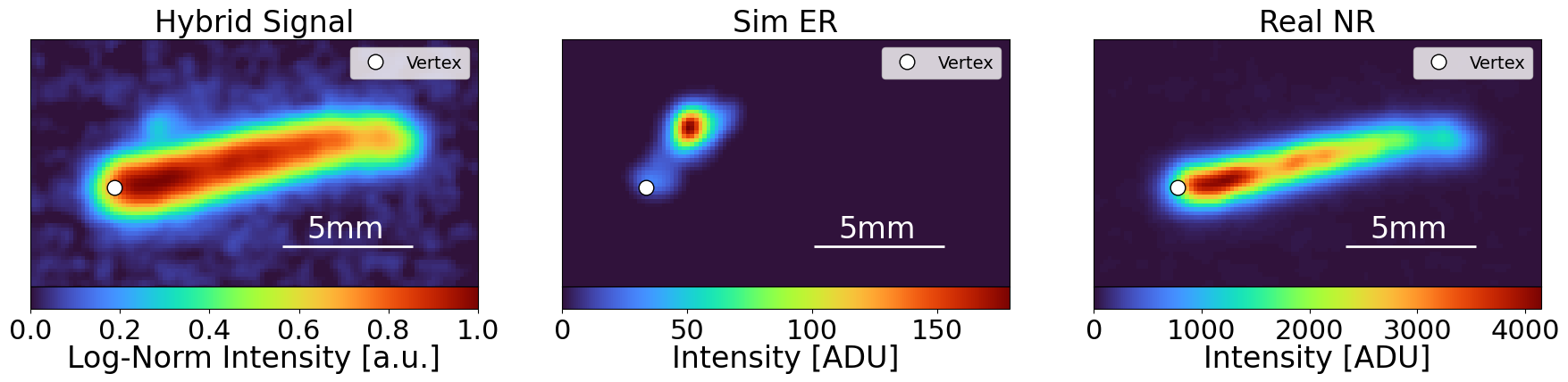}
\includegraphics[width=\linewidth]{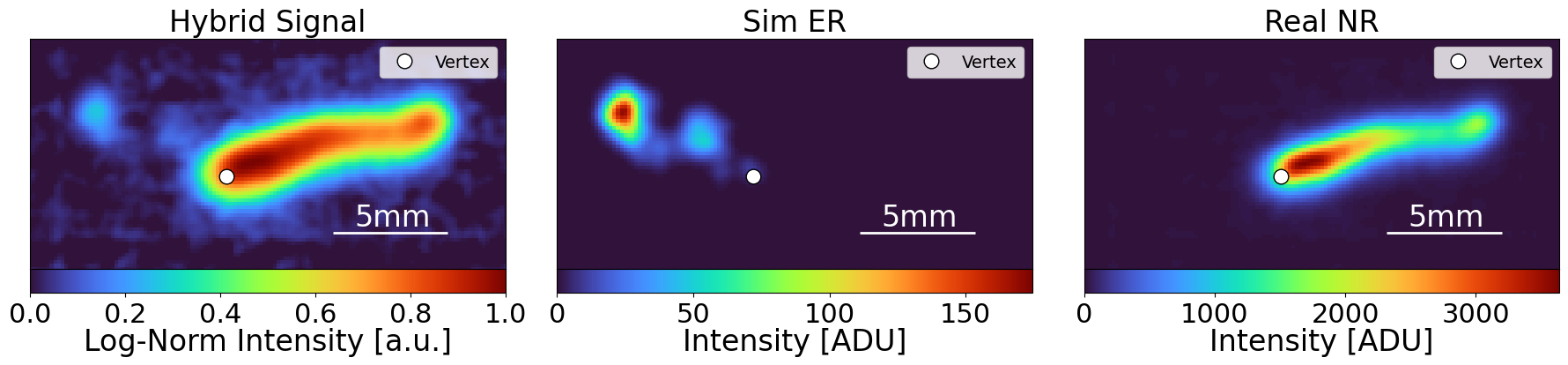}
\caption{Illustration showing the hybrid signal simulation construction of two events: one with a 5.8$\,$keV ER (top row) and the other with a 9.8$\,$keV ER (bottom row). For each event, the hybrid signal (left column) is constructed by stitching a simulated ER track (middle column) with a real, measured NR track (right column). The stitching point is the truth vertex position of the simulated ER and the estimated vertex of the real NR (both shown as white dots). \label{fig:hybrid}}
\end{figure}

Throughout the rest of this work, we evaluate OASIS's performance at reconstructing Migdal-effect ER signals that overlap with order(s)-of-magnitude brighter NRs. To do this, we simulate Migdal-effect signal events using the hybrid approach adopted in Ref.$\,$\cite{MIGDAL:2024alc}, where events are formed by stitching the truth vertex of a simulated ER track to the predicted vertex of a randomly-selected real, measured NR track.\footnote{In Ref.$\,$\cite{MIGDAL:2024alc}, we use the pixel of highest intensity in the NR track as a proxy for the its vertex. We now assign NR vertices more accurately using an algorithm based on curvilinear ridge detection$\,$\cite{Steger1998AnUD,Tilly:2023fhw} that will be presented in future work.} Figure~\ref{fig:hybrid} illustrates this construction for two example events with ER energies of 5.8$\,$keV (top row) and 9.8$\,$keV (bottom row). In this figure, the left column shows the hybrid signal (input to OASIS) while the middle and right columns show the ground-truth ER-only and NR-only intensity maps that serve as training targets. The simulation pipeline for generating ER tracks and the YOLO-based$\,$\cite{yolov8} pipeline used to define bounding box regions surrounding real NRs are described in Ref.$\,$\cite{MIGDAL:2024alc}. OASIS's task is to decompose each hybrid input image, $\mathcal{I}_\mathrm{in}(x,y)$ into $k=2$ output intensity maps: $\hat{\mathcal{I}}_\mathrm{ER}(x,y)$ representing the ER contribution, and $\hat{\mathcal{I}}_\mathrm{NR}(x,y)$, representing the NR contribution, to each pixel's intensity. For simplicity, the hybrid-simulated signal images formed in this work consist of exactly one ER and one NR. Further details about our image simulations, preprocessing, and labeling follow.

\subsection{Image Simulation, Preprocessing, and Labeling}
\label{subsec:preprocess}

The hybrid signal simulation approach offers several advantages for training and evaluation. First, using real NRs allows the network to learn authentic detector context while avoiding Sim2Real gaps$\,$\cite{9398246}; previous MIGDAL simulation studies$\,$\cite{Knights:2024wjh,MIGDAL:2024alc} have shown excellent pixel-level intensity and topological agreement between simulated and measured ER tracks, but comparable fidelity has not been achieved for NRs$\,$\cite{Edgeman:2026yhn}. Second, the hybrid approach produces clean ground truth labels: since detector noise is already present in the real NR images and the simulated ERs are noiseless, we obtain truth ER labels, $\mathcal{I}_\text{truth,ER}$, consisting purely of signal without double-counting noise. Finally, using simulated ERs eliminates ER vertex reconstruction uncertainties which are often larger than those of NRs. All together, the hybrid approach provides realistic signal simulation with truth labels that allow us to directly evaluate OASIS's segmentation and intensity reconstruction performance.


To make OASIS compatible out-of-the-box with MIGDAL's camera search software$\,$\cite{schueler_2024_12628437}, before training and evaluation, we preprocess each hybrid-simulated image using the same binning and smoothing procedure used for their search pipeline. Specifically, each image is $4\times 4$ binned and Gaussian smoothed with a $9\times 9$ kernel and $\sigma=4/3$, and all negative pixels are set to 0. The resulting images of dimension $n_x=512$ and $n_y=288$ are then normalized using a log-scaling procedure designed to compress the large dynamic range of pixel intensities while preserving faint ER features. For each input image $\mathcal{I}(x,y) \in \mathbb{R}^{512 \times 288}$, we determine a scale factor $s$ given by the 99th percentile of the pixel intensities of $\mathcal{I}$. Then, the following transformation is applied, yielding a scaled image with pixel values restricted to the unit interval:
\begin{equation}
\label{eq:logscale}
\mathcal{I}_\mathrm{in}(x,y) = \frac{\log_{10}\!\left(1 + \tfrac{\mathcal{I}}{s}\right)}{\max \log_{10}\!\left(1 + \tfrac{\mathcal{I}}{s}\right)}.
\end{equation}
This scaled image, $\mathcal{I}_\mathrm{in}(x,y)$, is what we input into our U--Net and the transformation from $\mathcal{I}$ to $\mathcal{I}_\mathrm{in}$  preserves contrast in the low-intensity regime, where ER tracks appear, and also prevents high-intensity NR pixels from completely dominating the dynamic range.  

Ground truth labels are constructed from the truth ER and NR portions of hybrid-simulated signal images. All together, event-data labels are stored with the following five quantities: 
\begin{enumerate}
    \item $\mathcal{I}_\mathrm{in}\in\mathbb{R}^{512\times 288}$
    \item$(\mathcal{I}_\mathrm{truth,ER},\mathcal{I}_\mathrm{truth,NR})\in\mathbb{R}^{2\times512\times288}$
    \item $s$ 
    \item $\max_{xy}(\mathcal{I}_\mathrm{in}(x,y))$
    \item A binary mask $B(x,y)$ defined to be 1 for the contents within the minimal enclosing bounding box of the event content of $\mathcal{I}_\mathrm{in}(x,y)$, and 0 elsewhere.
\end{enumerate} 
For each hybrid event, we define the bounding box region $B$ as the minimal enclosing bounding box of both the NR bounding box from our YOLO search pipeline$\,$\cite{MIGDAL:2024alc} and the truth boundary of the simulated ER\footnote{In real data where there is no knowledge of the truth ER boundary, $B$ could be defined as the NR bounding box determined by YOLO with an appropriate amount of additional padding to contain any expected Migdal ER in a given search region of interest. For a significant majority of hybrids with sub-6$\,$keV ERs evaluated in this work (those that tend to be most dominated by ER-NR overlap), it turns out the NR bounding box found by YOLO is equivalent to the hybrid bounding box, so no additional padding is needed. See the discussion in Section~\ref{subsec:ablation} for more details.}. Quantities 1., 2., and 5. are those relevant for optimizing our loss function and therefore the only relevant quantities for training, while 3. and 4. are used to map predictions back to physically meaningful units at the inference stage. At inference time, log-scaled channel predictions, $\hat{\mathcal{I}}_{c\mathrm{,log}}$, where $c\in{\{\mathrm{ER,NR}\}}$, can be mapped back to raw intensity units by inverting the transformation:
\begin{equation}
\hat{\mathcal{I}_c} = s \cdot \left( 10^{\hat{\mathcal{I}}_{c\mathrm{,log}}\cdot \max_{xy}(\mathcal{I}_\mathrm{in}(x,y))} - 1 \right),
\end{equation} 
ensuring that intensities can be reconstructed in physically meaningful units. 

\subsection{Model training}
OASIS's main task in this application is to extract the contents of an ER embedded within an NR in order to reconstruct characteristics of electron emission via the Migdal effect in order to compare with theory. It is therefore important for our model to be trained both on samples including ERs embedded within NRs (Migdal-effect signal events) and samples involving just NRs, in order for OASIS to learn the proper context necessary to avoid the hallucination of a non-existent ER signal. With this in mind, we utilize a mixed-class training sample consisting of 20,000 hybrid signal images and 20,000 NR-only images. Like the NRs used for the hybrid signal sample, the NR images used in the NR-only sample are also real NR tracks, but they are distinct from those used in the hybrid signal sample. Ground truth labels for the signal sample are 2-channel log-scaled (Equation~(\ref{eq:logscale})) intensity maps of the ER-only and NR-only channel of the input event, while for the NR-only sample, the ground truth ER channel is set to zero and is therefore an empty image.

To train our U--Net, we split both classes of our training sample into a 90$\%$--10$\%$ training--validation split. We train with a batch size of 16 and model weights are updated with the AdamW algorithm$\,$\cite{2017arXiv171105101L} using a learning rate of $2\times 10^{-4}$ and weight decay of $1\times 10^{-4}$. Training is terminated using early stopping$\,$\cite{estop} with a patience of 10 epochs, meaning that training is terminated when there are 10 successive epochs over which the validation loss does not improve. Ultimately, we perform 36 distinct training campaigns, each with different combinations of loss function weights, $w_c$ and $W_\mathrm{region}$, to systematically study the effects these weights have on OASIS's ability to reconstruct the faint ER channel. Each campaign was trained and evaluated across pairs of either Nvidia TITAN RTX or Nvidia A100 80GB graphics cards using PyTorch's$\,$\cite{paszke2019pytorch} Distributed Data Parallel platform$\,$\cite{li2020pytorch}. We do not perform any additional architecture or hyperparameter tuning, as our primary interest is quantifying the effects OASIS's novel loss function weights have on ER track reconstruction.

\section{Results applied to the MIGDAL experiment}
\label{sec:results}

We evaluate OASIS's performance on a test set consisting of 2,086 signal events (ER+NR) and 7,159 NR-only events. The test set was separated out from the training sample before training, so none of the test set events were used in training or validation.

\subsection{Loss function weighting ablation study}
\label{subsec:ablation}
\begin{figure*}
\includegraphics[width=.87\linewidth]{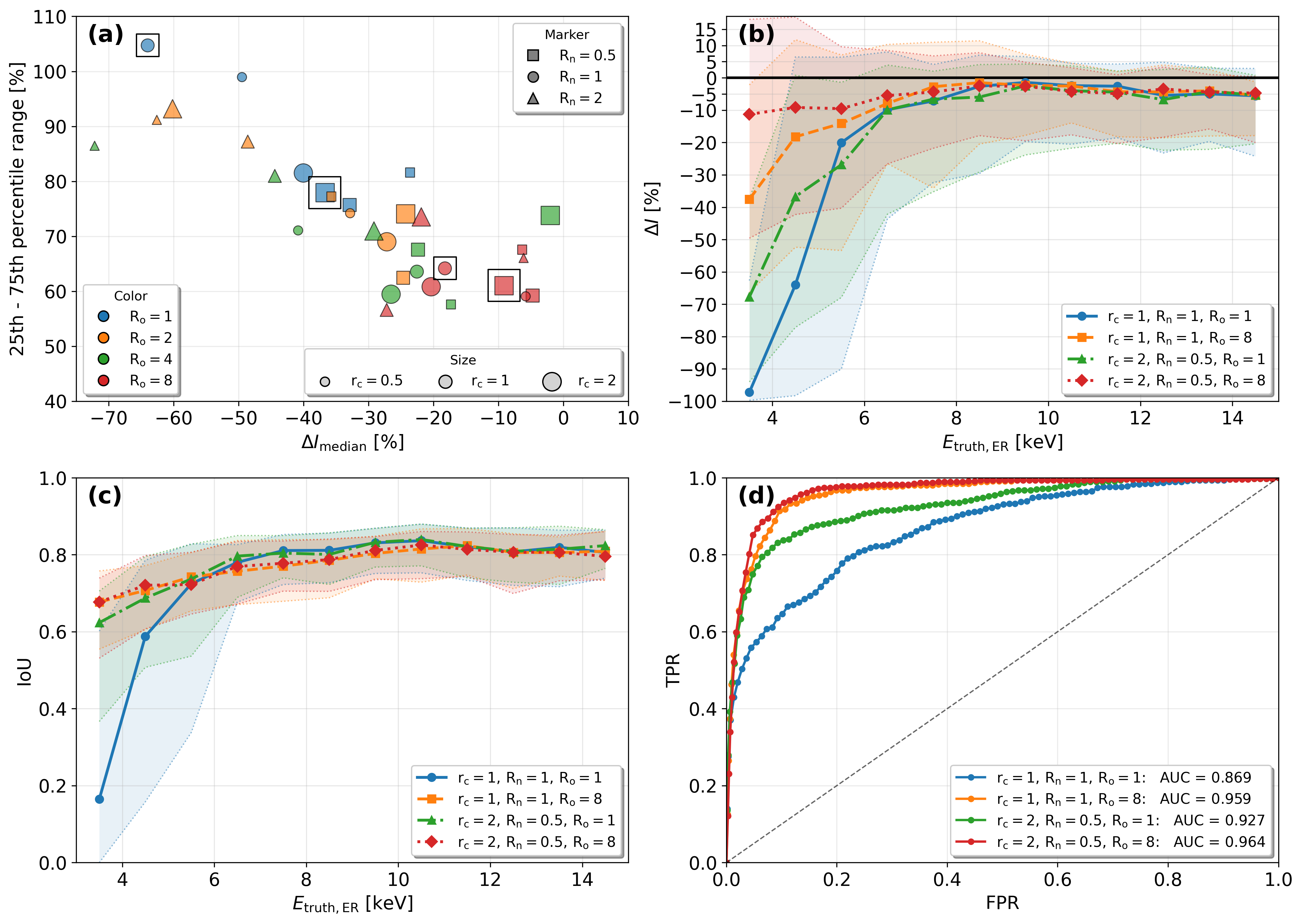}
\caption{Summary of ablation study results. (a) Spread (25th-75th percentile range) versus bias ($\Delta I_\mathrm{median}$) in ER-channel track intensity reconstruction in the signal event test sample (32 of the 36 weight combinations shown in this window). As described in the three legends, marker size, shape, and color denote values of $r_\mathrm{c}$, $R_\mathrm{n}$, and $R_\mathrm{o}$, respectively. (b) ER track intensity reconstruction bias (median shown by points) and spread (error bands showing 25th and 75th percentile ranges) versus truth ER energy for the four training campaigns shown within the black boxes in panel (a). (c) Same as (b) except with IoU vs. truth ER energy. (d) ROC curves for the same four training campaigns. True positive rates (TPR) are evaluated on the hybrid signal sample and are averaged over the 3-4$\,$keV, 4-5$\,$keV, and 5-6$\,$keV energy bins. False positive rates (FPR) are computed over the entire NR-only test set. \label{fig:ablation}}
\end{figure*}

To evaluate the effects OASIS's region and channel-specific loss function weights have on ER intensity and topological reconstruction, we perform a systematic ablation study across 36 training campaigns. The weights we vary are the ER and NR channel-specific weights $w_\mathrm{ER}$ and $w_\mathrm{NR}$, respectively, and the three region-specific weights $W_\mathrm{ER}$, $W_\mathrm{NR}$, and $W_\mathrm{overlap}$, consisting of ER-only, NR-only, and overlap region pixels, respectively. To reduce the dimension of weight combinations, we set baselines of $w_\mathrm{NR}=1$ and $W_\mathrm{ER}=1$ and then define the following three ratios to vary in our ablation studies
\begin{align}
\label{eq:ratios}
r_\mathrm{c}\equiv \frac{w_\mathrm{ER}}{w_\mathrm{NR}},\quad R_\mathrm{n}\equiv\frac{W_\mathrm{NR}}{W_\mathrm{ER}},\quad R_\mathrm{o}\equiv\frac{W_\mathrm{overlap}}{W_\mathrm{ER}}.
\end{align}
The results from evaluating OASIS models trained with all 36 unique combinations of $r_\mathrm{c}\in[0.5,1,2]$, $R_\mathrm{n}\in[0.5,1,2]$, and $R_\mathrm{o}\in[1,2,4,8]$, are summarized in Figure~\ref{fig:ablation} and described in detail below.

We are most interested in testing OASIS's ability to pixel-wise reconstruct the intensity map of the faint ERs buried within the significantly brighter NR. To this end, the metrics we reconstruct are (1) the bias and spread of the percent error of \textit{track} intensity reconstruction compared to truth, (2) the pixel intersection-over-union (IoU) of the reconstructed ER track compared to truth, and (3) the false positive rate of ERs reconstructed in the NR-only test sample. More explicitly, we define the track intensity reconstruction percent error as
\begin{align}
\Delta I\equiv \frac{\hat{I}_\mathrm{ER}-I_\mathrm{truth,ER}}{I_\mathrm{truth,ER}}\times 100\%,
\end{align}
where $I\equiv\sum_{xy}\mathcal{I}(x,y)$ represents track intensity, and IoU as
\begin{equation}
\mathrm{IoU}\equiv\frac{|\hat{\mathcal{I}}_\mathrm{ER}\cap\mathcal{I}_\mathrm{truth,ER}|}{|\hat{\mathcal{I}}_\mathrm{ER}\cup\mathcal{I}_\mathrm{truth,ER}|},
\end{equation}
where the numerator represents the number of shared pixels between the predicted and truth ER intensity maps, while the denominator represents their union. When computing IoU, we only include pixels with intensity of at least 1$\,$ADU.\footnote{Because we train with real-data NRs and simulated ERs, noise from the ORCA quest readout is only present in the NR channel. As a result, OASIS reconstructs ERs without noise, hence the low 1$\,$ADU threshold for IoU masking.} We report the bias and spread of both $\Delta I$ and IoU as their median and 25th-75th percentile range, respectively. Taken together, these two quantities characterize both ER intensity attribution and topological reconstruction performance.

Figure~\ref{fig:ablation}(a) summarizes intensity attribution bias and spread across all ablations for ER tracks in the 4-5$\,$keV energy bin. We intentionally investigate a low energy ER bin because the region of pixel overlap between the ER and NR, on average, encompasses a larger proportion of low energy ER tracks compared to high energy tracks. The effects of overlap-weighted training should therefore be most pronounced for low energy ER tracks, which is confirmed in panels (b) and (c) of Figure~\ref{fig:ablation}. Independent of $R_\mathrm{n}$ and $r_\mathrm{c}$, it is immediately apparent from Figure~\ref{fig:ablation}(a), and quantified in Table~\ref{tab:ablation_results}, that increasing $R_\mathrm{o}$ improves both bias and spread OASIS's track intensity reconstruction error.

Narrowing in on the four ablations shown in Figure~\ref{fig:ablation}(b)-(c) allows for a holistic comparison across all ER energies between the case of unweighted segmentation regression, the usage of weights prioritizing ER reconstruction, and the usage of weights prioritizing overlap-region reconstruction. Comparing the traces with blue circles and green triangles, we observe an improvement in intensity reconstruction and a significant improvement in topological IoU score in the 3-4$\,$keV and 4-5$\,$keV ER energy bins. This result shows that applying channel and region-specific weights that explicitly prioritize ER reconstruction over NR reconstruction in OASIS's training improves performance over unweighted training in the ERs that overlap most with NRs. OASIS's overlap region weights, however, yield more significant improvements in overlap-dominated ER reconstruction than weights prioritizing ER reconstruction, as is shown by the trace with orange squares. Finally, the trace with red diamonds includes weights that prioritize both ER reconstruction as well as overlap-targeted weights, yielding the best intensity reconstruction performance for low energy ERs where attribution is most challenging. For example, in the 4-5$\,$keV bin, where the ER-NR overlap region tends to comprise a large proportion of the ER track, OASIS trained with $(r_\mathrm{c},R_\mathrm{n},R_\mathrm{o})=(2,0.5,8)$ achieves $\Delta I_\mathrm{median}=-9.15\%$ and $\mathrm{IoU}_\mathrm{median}=0.72$, compared to $\Delta I_\mathrm{median}=-64.0\%$ and $\mathrm{IoU}_\mathrm{median}=0.59$ for unweighted segmentation regression.\footnote{For clarity, unweighted segmentation regression has $(r_\mathrm{c},R_\mathrm{n},R_\mathrm{o})=(1,1,1)$ which is equivalent to the segmentation-regression loss term optimizing the mean absolute error for each object.}

\begin{table}[htbp]
\centering
\begin{tabular}{ccc}
\toprule
$R_\mathrm{o}$ & Bias (\%) & Spread (\%) \\
\midrule
1  & $-41.1$ & 86.8 \\
2    & $-39.5$ & 78.6 \\
4     & $-30.8$ & 70.2 \\
8    & $\mathbf{-13.3}$ & \textbf{63.1} \\
\bottomrule
\end{tabular}
\caption{Bias ($\Delta I_\mathrm{median}$) and spread (25th-75th percentile range) of track intensity reconstruction in the 4-5$\,$keV ER-energy bin of the hybrid signal test set, averaged across all ablations of $r_\mathrm{c}$ and $R_\mathrm{n}$ for each value of overlap-weight ratio $R_\mathrm{o}$. Increasing overlap penalty weights in the OASIS's loss function significantly improves the intensity reconstruction performance of low energy ER tracks that are proportionately dominated by regions of pixel overlap, compared to higher energy tracks.}
\label{tab:ablation_results}
\end{table}

From these results, it is clear that OASIS's use of region and channel weights significantly improves both pixel-level intensity attribution and topological reconstruction of low energy ER tracks. This improvement in the low-energy regime is particularly important because the Migdal effect cross section increases exponentially with decreasing ER energy$\,$\cite{Cox:2022ekg}, making accurate reconstruction in the 3-6$\,$keV range essential for achieving sufficient statistics to test theoretical predictions. We additionally note that around $80\%$ of events in the hybrid signal sample with sub-6$\,$keV ERs are evaluated by OASIS with bounding box mask regions that are identical to YOLO's NR bounding box prediction. Therefore, the inputs corresponding to a significant majority of low ER-energy events do not utilize any information at inference that would not be available in a real-data pipeline. This gives confidence in the potential for OASIS's overlap region reconstruction performance gains to generalize to real data.

The final test of this ablation study compares OASIS's ER construction performance on both the hybrid signal and NR-only test sets to evaluate true and false positive detection rates, respectively. By design, OASIS should always output a nonzero amount of ER signal, even if the ER isn't present, so modest track intensity thresholds must be applied to distinguish positive versus negative ER detections. Figure~\ref{fig:ablation}(d) shows ROC curves derived from varying the reconstructed intensity thresholds of the ER-channel from which a positive or negative ER detection is decided. These ROC curves are generated for the four ablations we've narrowed our focus to, with true positive rates (TPR) averaged over the 3-4$\,$keV, 4-5$\,$keV, and 5-6$\,$keV truth ER energy bins of the hybrid signal samples, false positive rates (FPR) evaluated on the NR-only sample. Using the area under the ROC curve (AUC) as our performance metric, we observe the same relative ranking in performance as Figure~\ref{fig:ablation}(b) and (c). In other words, the overlap-region weights introduced in OASIS are the single most important loss function weight for improving TPR at fixed FPR, and the training campaign that weights both ER reconstruction and overlap region reconstruction performs best.

In summary, our systematic ablation study shows that ER-targeted- and especially overlap-region targeted- loss function weights improve ER intensity and topological reconstruction, and also improve true positive detection rates at fixed false positive rates. Given that the $(r_\mathrm{c},R_\mathrm{n},R_\mathrm{o})=(2,0.5,8)$ training campaign is the best performing for all metrics evaluated in detail in this ablation study, we spend the remainder of this section presenting more detailed results from this campaign. Before further quantitative studies, we first examine this training campaign's reconstruction performance on individual test set events to illustrate typical reconstruction quality and failure modes.

\subsection{Event-level reconstruction}
\begin{figure*}
\includegraphics[width=\linewidth]{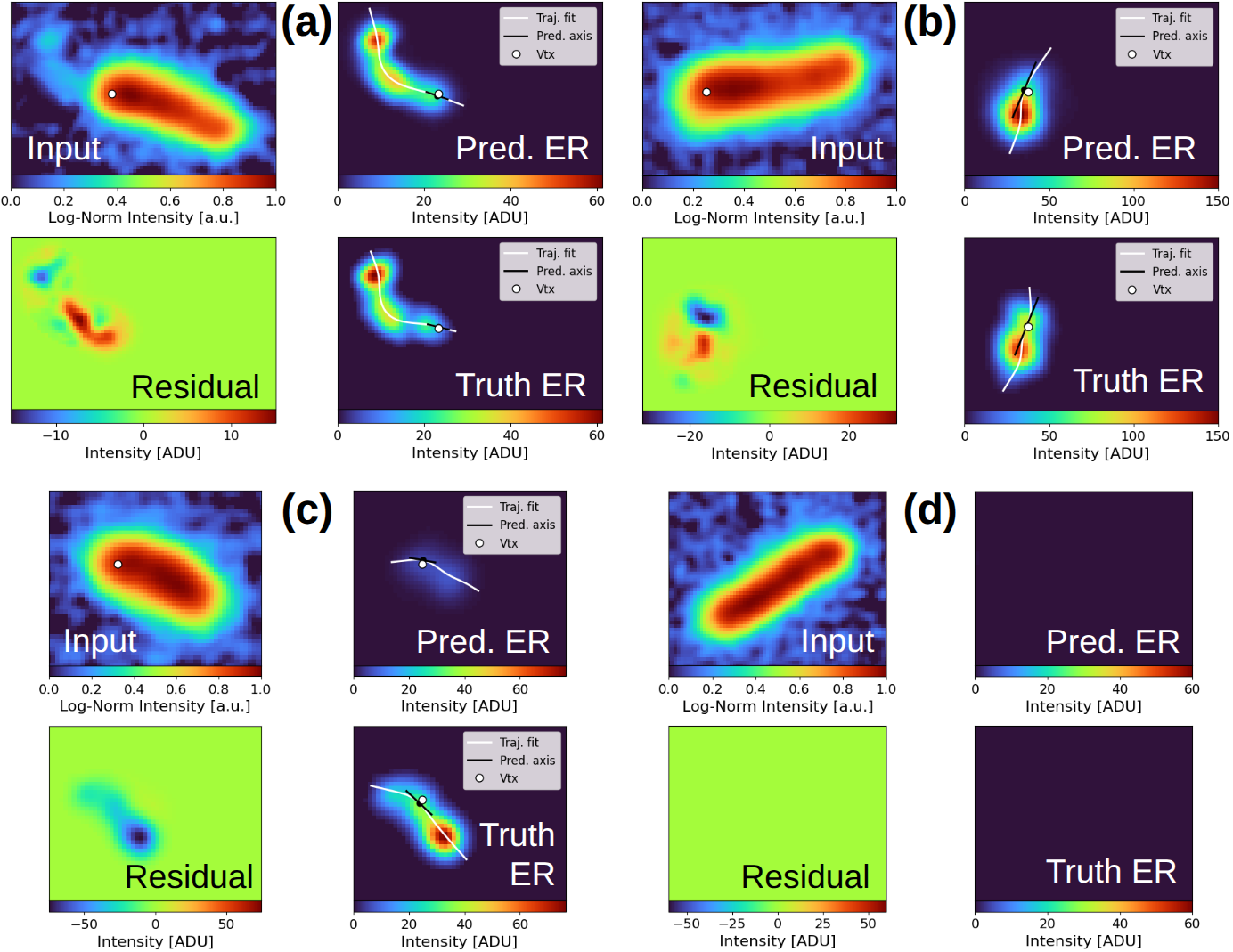}
\caption{Four test set examples (from the $(r_\mathrm{c},R_\mathrm{n},R_\mathrm{o})=(2,0.5,8)$ training campaign) comparing OASIS's ER reconstruction performance to truth. In panels (a)-(c) both the truth and predicted ERs have principal curves shown in white with estimated directional axes shown in black. Panel (d) is an NR-only input event. Predicted ER and truth ER energies for each panel are: (a) $\hat{E}_\mathrm{ER}=$ 5.9$\,$keV,  $E_\mathrm{truth,ER}=$ 5.2$\,$keV. (b) $\hat{E}_\mathrm{ER}=$ 5.4$\,$keV,  $E_\mathrm{truth,ER}=$ 5.2$\,$keV. (c) $\hat{E}_\mathrm{ER}=$ 0.7$\,$keV,  $E_\mathrm{truth,ER}=$ 5.7$\,$keV. (d) $\hat{E}_\mathrm{ER}=$ 12$\,$eV,  $E_\mathrm{truth,ER}=$ 0$\,$eV. For panels (a)-(c) $\hat{E}_\mathrm{ER}\equiv\frac{\hat{I}_\mathrm{ER}}{I_\mathrm{truth,ER}}E_{\mathrm{truth,ER}}$, and for panel (d) $\hat{E}_\mathrm{ER}$ is computed from scaling $\hat{I}_\mathrm{ER}$ relative to the reference intensity of a 5.9$\,$keV ER track.\label{fig:output_examples}}
\end{figure*}
\begin{figure}
\includegraphics[width=0.31\linewidth]{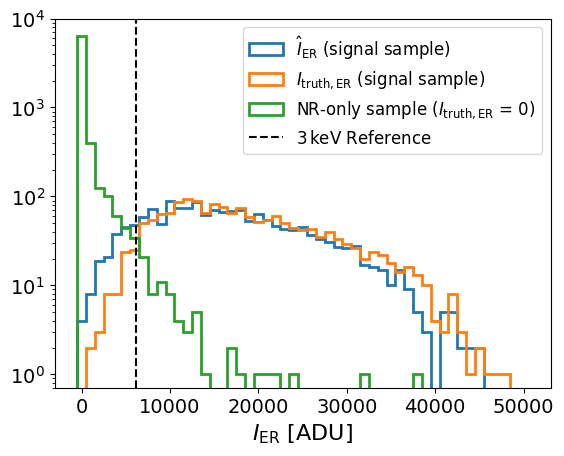}
\includegraphics[width=.69\linewidth]{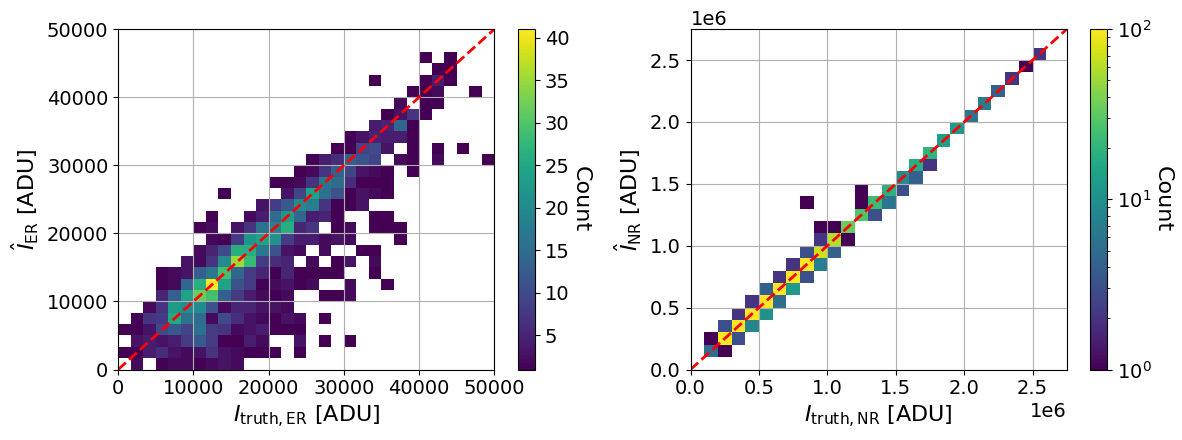}
\caption{Left: Histograms of reconstructed and truth ER intensities for the test set hybrid signal sample (blue and orange, respectively), and reconstructed ER intensities for the NR-only sample (green). The black vertical dashed line shows the average intensity of a 3$\,$keV ER. Middle: Model's predicted ER track intensity ($\hat{I}_\mathrm{ER}$) versus truth ER track intensity ($I_\mathrm{truth,ER}$) for all test set ERs satisfying $4\leq E_\mathrm{truth,ER}\leq 15$$\,$keV. The red line shows equality between prediction and truth illustrating that OASIS'S ER track energy reconstructions are strongly correlated with truth. Right: Same as middle panel but for NR intensity reconstruction. All predictions in this figure come from OASIS trained with $(r_\mathrm{c},R_\mathrm{n},R_\mathrm{o})=(2,0.5,8)$.\label{fig:ER_reco}}
\end{figure}

Figure~\ref{fig:output_examples} illustrates OASIS's performance on four examples from the test set. For the shown input image, panels (a)-(d) of this figure show comparisons between OASIS's predicted ER and the ground truth ER associated with the input image. In panels (a)-(c), we compare (1) OASIS's \textit{track} intensity prediction with truth, $\Delta I$, and (2) assess the consistency of axial angular reconstruction by computing the angle between axial directions -- determined by applying the same angle-finding algorithm -- computed on the predicted and truth ER intensity maps, $\Delta\phi\in[0^\circ,90^\circ]$. We call the $\Delta \phi$ performance metric \textit{angular consistency}, as it is a measure of similarity between reconstructed angles on both the ground truth ER image and the ER image reconstructed by OASIS. We note the distinction between this and angular resolution, which compares a reconstructed track angle with the true initial angle of the ER. OASIS's goal is to faithfully separate an ER from an overlapping NR, not to reconstruct the true ER direction, so angular consistency is a more practical performance measure for OASIS's ER topology reconstruction. Details of the angular reconstruction algorithm from which we compute angular consistencies are discussed in Appendix~\ref{sec:A2}. 

Panel (a) of Figure~\ref{fig:output_examples} shows a typical example of an event where the ER track is long enough and directed away from the NR so that a significant portion of its topology lies outside of the NR. In this case, OASIS does an excellent job reconstructing both the intensity ($\Delta I=+13\%$) and capturing directional information near the ER's vertex ($\Delta\phi=5.7^{\circ}$). The example in panel (b) is noteworthy because the ER is hardly visible by eye in the input image, yet OASIS still successfully extracts the ER ($\Delta I=+4\%$, $\Delta\phi=0.9^\circ$). Panel (c) illustrates a typical example where our model does not perform well. In this event, the truth ER travels inward toward the brightest regions of the NR. Despite weighting the ER-NR overlap region most heavily in our network training, OASIS predicts too little overall intensity. For this event, we find $\Delta I=-88\%$, and $\Delta\phi=33^\circ$, hinting at a correlation between energy and angular reconstruction performance, which will be discussed later. Finally, panel (d) highlights typical reconstruction performance for an NR-only input event. In this case OASIS predicts about 12$\,$eV of visible energy which is effectively zero compared to the keV-scale ERs of interest in this study, so we treat this example as OASIS successfully determining that there was no ER in this input image. Having illustrated event-level performance, we now assess OASIS's intensity reconstruction across the full test set to quantify systematic trends and overall accuracy.

\subsection{Intensity reconstruction performance}
The green histogram in the left panel of Figure \ref{fig:ER_reco} shows the overall distribution of reconstructed ER track intensities for the NR-only sample. Noting the logarithmic scale, strong peaking near 0, and that the lowest truth ER energy in the hybrid signal sample is 3$\,$keV, we observe that OASIS correctly predicts virtually no ER signal in the vast majority of NR-only events. By contrast, in the hybrid signal sample, non-negligible amounts of ER signal are nearly always detected.

The middle panel of Figure \ref{fig:ER_reco} shows the correlation between OASIS's ER track intensity predictions with truth for events satisfying MIGDAL's search region of interest of $4\leq E_\mathrm{truth,ER}\leq 15$$\,$keV. We make the following observations from this distribution. First, the red line indicates perfect performance ($\hat{I}_\mathrm{ER}=I_\mathrm{truth,ER}$) and much of the data is tightly centered around this line.
Second, while the mode of the $\hat{I}_\mathrm{ER}$ distribution at most slices of $I_\mathrm{truth,ER}$ aligns with perfect performance, at lower intensities, the model is more likely to under-predict the ER's intensity, which was also shown in Figure~\ref{fig:ablation}(b).

Finally, the right panel shows the correlation between OASIS's NR track intensity predictions with truth in the hybrid signal sample. Given that NR signal intensities overwhelmingly dominate over ER signals in all images, we expect near perfect NR intensity reconstruction, which is what we observe.

Beyond intensity reconstruction, accurate angular information is essential for comparing observations with theoretical Migdal emission models. We therefore assess OASIS's angular reconstruction consistency using principal curve analysis.

\subsection{Angular consistency performance}
\begin{figure}[htbp]
\includegraphics[width=0.45\linewidth]{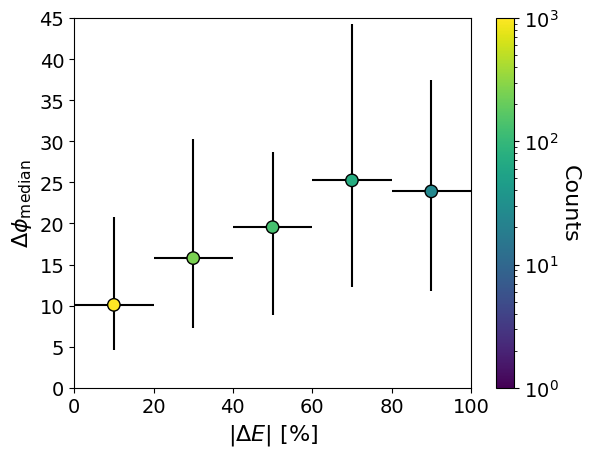}
\hspace{0.05\linewidth}
\includegraphics[width=0.44\linewidth]{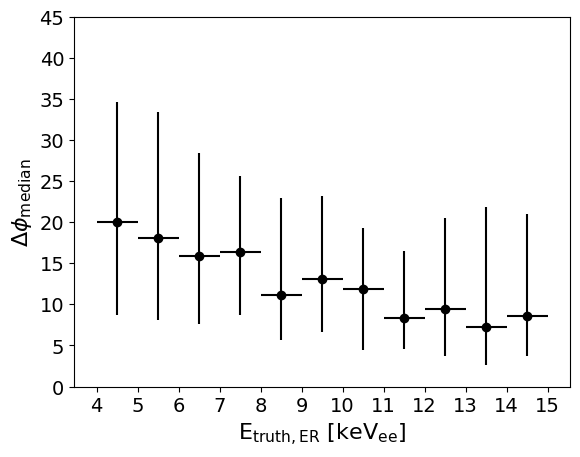}
\caption{Angular reconstruction performance summary for the $(r_\mathrm{c},R_\mathrm{n},R_\mathrm{o})=(2,0.5,8)$ training campaign. Left: Median (points) and inter-quartile range (error bars) of angular consistency $\Delta\phi$ versus $|\Delta I|$. Right: Median (points) and inter-quartile range (error bars) or $\Delta\phi$ versus truth ER energy.\label{fig:ang_consistency_err}}
\end{figure}
To assess OASIS's angular reconstruction performance, we fit principal curves (details in Appendix~\ref{sec:A2}) to the predicted and truth ER intensity maps for events in the signal sample of the test set satisfying $4\leq E_\mathrm{truth,ER}\leq 15$$\,$keV. Of these 1,750 events, principal curves were successfully fit on 1,571 of them (90$\%$). Our interest in this study is assessing the topological similarity between OASIS's predictions and truth in a way that captures relevant detector performance metrics. We therefore use the axial angle between the reconstructed axis of the truth and predicted track topologies, $\Delta\phi\in[0,90^\circ]$, as our heuristic for topological similarity. The left panel of Figure~\ref{fig:ang_consistency_err} shows the median (points) and 25th-75th percentile values (errorbars) of $\Delta\phi$ versus the absolute percentage difference in reconstructed intensity versus truth, $|\Delta I|$. It is immediately clear that angular reconstruction performance is strongly correlated with intensity reconstruction performance. This tells us that in general, OASIS's ER intensity reconstructions are spatially correlated and tend toward the ``correct answer" both in terms of energy and angle. While this is expected because we designed our loss function to simultaneously optimize event segmentation and energy reconstruction, this result confirms this optimization in a physically meaningful way.

The right panel of Figure~\ref{fig:ang_consistency_err} shows the median and 25th-75th percentile ranges of angular consistency versus truth ER energy. While we do not compute angular resolutions with respect to true recoil directions, the fact that median angular reconstruction consistencies are within $20^\circ$ over all energies $\geq$4$\,$keV is a promising first step toward demonstrating sensitivity in measuring angular distributions of Migdal ER emissions. Indeed, if the angular resolution of a detector for isolated ERs is determined by some algorithm, then the angular resolution of ERs emitted via the Migdal effect can be estimated by adding $\Delta\phi$ -- determined by the same algorithm -- in quadrature to the detector's angular resolution.

\section{Discussion}
\label{sec:conclusion}

The regions where objects overlap are precisely where attribution is most ambiguous and informative. Yet these critical regions are typically given no special consideration in segmentation-regression training. OASIS addresses this by introducing explicit overlap-region weighting into its segmentation-regression loss function. By assigning both channel and region-specific weights to areas where multiple objects intersect, the framework directs training attention toward the regions where accurate attribution is most critical.

When tasked with reconstructing the MIGDAL experiment's faint ER signal -- one that is often buried within an order(s)-of-magnitude brighter NR signal -- we demonstrated that weighting OASIS's loss function to prioritize regions of pixel overlap between the ER and NR yields substantial improvement in intensity attribution, event topology reconstruction, and true positive detections of low energy ERs at fixed false positive rates. In particular, in the 4-5$\,$keV ER energy regime where ER-NR overlap tends to dominate the ER signal, averaging over all combinations of weights targeting ER/NR reconstruction, $r_\mathrm{c}$ and $R_\mathrm{n}$, at fixed overlap weight ratios, $R_o$ (Eq.~(\ref{eq:ratios})), we find $\Delta I_\mathrm{median}$ to improve from $-41.1\%$ at $R_\mathrm{o}=1$ to $-13.3\%$ at $R_\mathrm{o}=8$ (Table~\ref{tab:ablation_results}). Furthermore, compared to unweighted training, the addition of OASIS's novel overlap region-targeted weight in its loss function is the single most important adjustment in improving the reconstruction of low energy ER tracks across all metrics considered, as evidenced by comparing the blue and orange traces in Figure~\ref{fig:ablation}(b)-(d). Linking OASIS's topological reconstruction performance to more physically meaningful quantities, with the choice of $(r_\mathrm{c},R_\mathrm{n},R_\mathrm{o})=(2,0.5,8)$, we observe angular consistencies derived from principal curve fits to within 20$^\circ$, down to ER energies of 4$\,$keV. Given the exponentially rising Migdal emission probability with decreasing ER energy, achieving sensitivity to reconstruct ER energies and angles in this regime is essential for achieving sufficient statistics to meaningfully test theoretical predictions. OASIS's angular sensitivity in this regime shows promise toward the current generation MIGDAL detector's capability of testing theoretical models of Migdal effect angular production, which we will report on in future work.

Beyond MIGDAL, the principles underlying OASIS extend to other scientific imaging domains where overlapping objects must be separated. In astronomy there are publicly accessible datasets like the Galaxy Zoo DECaLS database$\,$\cite{2022MNRAS.509.3966W}, which contains hundreds of thousands of images that are pre-annotated for the purpose of training models for tasks like ``deblending" overlapping galaxy images in crowded fields. The fact that such datasets are publicly available and that intensity attribution in overlapping regions is an essential component of galaxy deblending makes it an attractive avenue to apply OASIS. The OASIS framework can be readily extended to support this, as well as other diverse applications: $n$-channel inputs enable multi-band astronomical imaging, while full 3D voxel support (3 spatial dimensions + intensity) would enable application to volumetric medical imaging and upgraded particle detectors with complete 3D reconstruction capabilities. We are additionally exploring the possibility of extending OASIS's overlap-aware loss weights to 1D waveform data, where the task is to sample-wise reconstruct a physical signal channel and a background channel. Applying OASIS to such a task could potentially enable weak signal extraction in regimes where traditional thresholding fails. All together, OASIS's ability to recover information from challenging regions of object-overlap gives it the potential to improve scientific image analyses across disciplines.

\section*{Data Availability Statement}
All code supporting this study is openly available at \url{https://github.com/jschuel/OASIS}.

\ack{This work has been supported by the UKRI’s Science $\&\,$Technology Facilities Council through the Xenon Futures R$\,\&\,$D programme (awards ST/T005823/1, ST/T005882/1, ST/V001833/1, ST/V001876/1), Consolidated Grants (ST/W000636/1, ST/X006042/1, ST/T000759/1, ST/W000652/1, ST/S000860/1, ST/X005976/1), and TM’s and LM’s PhD scholarships (ST/T505894/1, ST/X508913/1); by the U.S. Department of Energy, Office of Science, Office of High Energy Physics, under Award Number DE-SC0022357; by the U.S. National Science Foundation under Award number 2209307; ET acknowledges the Graduate Instrumentation Research Award funded by the U.S. Department of Energy, Office of Science, Office of High Energy Physics; by the Portuguese Foundation for Science and Technology (FCT) under award number PTDC/FIS-PAR/2831/2020; and by the European Union’s Horizon 2020 research and innovation programme under the Marie Skłodowska-Curie grant agreement No. 841261 (DarkSphere) and No. 101026519 (GaGARin). ELA acknowledges the support from Spanish grant CA3/RSUE/2021-00827, funded by Ministerio de Universidades, Plan de Recuperacion, Transformacion y Resiliencia, and Universidad Autonoma de Madrid. KN acknowledges support by the Deutsche Forschungsgemeinschaft (DFG, German Research Foundation) under Germany's Excellence Strategy -- EXC 2121 "Quantum Universe“ -- 390833306. We are grateful to the Particle Physics Department at RAL for significant additional support which made this project possible. Thanks are also due to the CERN RD51 collaboration for their support through Common Project funds, hardware tests and training, and useful discussions. We would also like to thank the ISIS facility for technical assistance and for hosting this experiment. We thank Master’s students M.~Handley (Cambridge), R.~Hafeji (Surrey), and G.~Buzzard (Liverpool).}

\appendix

\section{Appendix A: Directional reconstruction algorithm}
\label{sec:A2}
When assessing the directional reconstruction performance of OASIS, our aim is to compare angles reconstructed in a consistent way between our model's predicted ER image and the truth ER image. To this aim, any angular reconstruction algorithm can be used to estimate $\Delta\phi$. Because electron recoil trajectories are varied, and often non-linear, we opt to evaluate $\Delta\phi$ using an algorithm that captures track curvature. The principal curve fitting technique$\,$\cite{hastie} does just this and has a readily-adaptable implementation$\,$\cite{princurve} in the R programming language$\,$\cite{R}. Using this implementation, our angular reconstruction procedure for a given ER intensity map is as follows:

\begin{enumerate}
    \item Sample 20,000 coordinates drawn from the non-zero bin-centers of the input intensity map, weighted by the log-scale intensity from Equation~(\ref{eq:logscale}). We do this because the principal curve fitting implementation does not support weights as inputs, but does properly weight duplicate coordinates.
    \item Generate a principal curve from the 20,000 input coordinates. We apply a convergence threshold of 0.25$\,$\cite{princurve} and interpolate the curve with a smoothed spline.
    \item Construct a principal axis as the tangent line to the location along the principal curve that has the closest Euclidean distance to the truth ER vertex.
\end{enumerate}
After performing steps 1.--3. on both OASIS's predicted ER intensity map and on the truth ER intensity map, we compute $\Delta\phi$ as the angle corresponding to the dot product between the two normalized axes, folded to the domain $\Delta\phi\in[0^\circ,90^\circ]$. 

Our principal axis construction is idealized in the sense that we utilize the truth ER's vertex position from simulation. ER vertices are challenging to accurately pinpoint, however since the ER and NR share a vertex in the Migdal effect, we can construct principal axes for Migdal ERs at our estimate of the NR vertex position. Our collaboration recently developed an algorithm based on our previous work on curvilinear ridge detection$\,$\cite{Steger1998AnUD,Tilly:2023fhw} that correctly reconstructs NR vertices to sub-pixel-level precision (publication forthcoming). Assuming this vertex placement algorithm behaves similarly on OASIS's predicted \textit{NR} outputs to how it does on raw NRs, then we expect ER directions evaluated at the point on a principal curve nearest to the predicted NR vertex to be similar to those evaluated nearest to the truth ER vertex.
\bibliographystyle{unsrt}
\bibliography{apssamp}

\end{document}